\documentclass[HFI]{baltzer}
\begin{document}
\begin{frontmatter}
\title{Continuous measurements in quantum systems}
\author[A]{Carlo Presilla} and 
\author[B]{Ubaldo Tambini}
\address[A]{Dipartimento di Fisica, Universit\`a di Roma ``La Sapienza,''
and INFN, Sezione di Roma,\\ Piazzale A. Moro 2, Roma, Italy 00185}
\address[B]{Dipartimento di Fisica, Universit\`a di Ferrara,
and INFN, Sezione di Ferrara,\\ Via Paradiso 12, Ferrara, Italy 44100}
\runningauthor{Presilla,Tambini}
\runningtitle{Continuous measurements in quantum systems}
\begin{abstract}
During a continuous measurement, quantum systems can be described by a 
stochastic Schr\"odinger equation which, in the appropriate limit, 
reproduces the von Neumann wave-function collapse.
The average behavior on the ensemble of all measurement results is described
by a master equation obtained from a general model of measurement apparatus 
consisting of an infinite set of degrees of freedom linearly interacting with 
the measured system and in contact with a reservoir at high temperature.
\end{abstract}
\end{frontmatter}

In ordinary quantum mechanics measurements are taken into account 
by postulating the wave function collapse \cite{VONNEUM}.  
This approach has the unpleasant feature of introducing 
an extra assumption in the theory and deals only with instantaneous 
and perfect measurements. 
It is more realistic and satisfactory to recognize that a measured 
system is not isolated but in interaction with a measurement apparatus.
Of course, a detailed description of a particular 
measurement apparatus would have limited utility.
One needs an effective quantum equation which, while reproducing
the main features of a measurement process, does not  explicitly contain 
the degrees of freedom of the meter. 

During the past 20 years several approaches have been attempted in the
above mentioned direction.
It turns out that, apart from a difference in language and points of view,
a mathematical equivalence can be established among all them \cite{MQM}.
Here, we briefly review some aspects of the corresponding modified quantum 
mechanics which includes the effects of a continuous measurement. 

Let's consider a system initially prepared in the state $| \psi(0) \rangle$. 
During the measurement of an observable corresponding to the operator $\hat A$
this state should evolve, preserving its norm, into a state close to an 
eigenstate of $\hat A$. 
The evolution must have a stochastic character since different measurements 
with the same initial state $| \psi(0) \rangle$ may give different results.
All these requirements are satisfied by the
nonlinear quantum state diffusion equation, proposed in 
\cite{DIOSI1,BELAVKIN1,GISIN2},   
\begin{eqnarray}
d | \psi_{[\xi]}(t) \rangle &=& 
\left\{ - {i \over \hbar} \hat H 
- {1 \over 2} \kappa \left[ \hat A - 
\langle \psi_{[\xi]}(t) | \hat A |\psi_{[\xi]}(t) \rangle 
\right]^2 \right\}~ |\psi_{[\xi]}(t) \rangle ~dt 
\nonumber \\ &&
+\sqrt{\kappa} \left[ \hat A - 
\langle \psi_{[\xi]}(t) | \hat A |\psi_{[\xi]}(t) \rangle  \right] 
~|\psi_{[\xi]}(t) \rangle~\xi(t)dt\,.  
\label{STOCEQ}
\end{eqnarray}
The stochastic process $\xi(t)$ can be chosen as a real white noise 
\begin{equation}
\int d[\xi]~\xi(t) = 0, ~~~~~~~~~~ 
\int d[\xi]~ \xi(t)~\xi(s) = \delta(t-s). 
\end{equation}
with respect to the functional integration measure
\begin{eqnarray}
d[\xi] &=&  \lim_{N \to \infty}
\prod_{n=1}^{N} d\xi^{(n)} \sqrt{t \over 2 \pi N} 
\exp \left( -{ {t \xi^{(n)}}^2 \over 2N} \right).
\label{XIMEASURE}
\end{eqnarray}
The constant $\kappa$  measures the strength of the coupling with the
(infinite) degrees of freedom of the measurement apparatus represented by 
$\xi(t)$. \\
Equation (\ref{STOCEQ}) reduces to the ordinary Schr\"odinger equation 
for $\kappa=0$. 

The stochastic process 
$\langle \psi_{[\xi]}(t) | \hat A |\psi_{[\xi]}(t) \rangle$
is the result of a single continuous measurement.
If many measurements are repeated with the measured system always
prepared in the same initial state $|\psi(0) \rangle$, we define 
an average result and an associated variance
by averaging over the Gaussian stochastic process $\xi(t)$
\begin{equation}
\overline{a(t)} = 
\int d[\xi] ~ \langle \psi_{[\xi]}(t) | \hat A |\psi_{[\xi]}(t) \rangle,  
\label{AVEXI}
\end{equation}
\begin{equation}
\Delta a(t)^2 =
\int d[\xi] ~ \langle \psi_{[\xi]}(t) | 
\left[ \hat A - \overline{a(t)} \right]^2 |\psi_{[\xi]}(t) \rangle.
\label{VARXI}
\end{equation}
These and other averaged quantities can be evaluated in a more direct 
way by considering a dynamical equation for the averaged density matrix
operator 
\begin{equation}
\hat \rho(t) = \int d[\xi] ~
| \psi_{[\xi]}(t) \rangle \langle \psi_{[\xi]} (t) |.
\end{equation}
By using Ito algebra and Eq. (\ref{STOCEQ}), we get 
\begin{eqnarray}
{d \over dt} \hat \rho(t) = 
- {i \over \hbar} \left[ \hat H, \hat \rho(t) \right] 
- {1 \over 2} \kappa
\left[ \hat A, \left[ \hat A, \hat \rho(t) \right] \right]
\label{LINDBEQ}
\end{eqnarray}
which reduces to the standard quantum master equation for $\kappa=0$.
Equation (\ref{LINDBEQ}) is of Lindblad class and this ensures the positivity 
of $\hat \rho(t)$ \cite{GORINI,LINDBLAD}.
Since no realizations of the stochastic process $\xi(t)$ have now to be 
selected, we call {\it nonselective measurement} the process described by 
Eq. (\ref{LINDBEQ}). 
{\it Selective measurements} are those described by Eq. (\ref{STOCEQ}).
According to the usual quantum rules, we finally write
\begin{equation}
\overline{a(t)} = {\rm Tr} \left[ \hat{A} \hat\rho(t) \right]\,, 
\label{AVENS}
\end{equation}
\begin{equation}
\Delta a(t)^2 =
{\rm Tr} \left[ \left( \hat A -\overline{a(t)} \right)^2 \hat \rho(t) 
\right]
\label{VAR}
\end{equation}
and similarly for other averaged quantities.

The usual von Neumann collapse can be recovered
in the case of an impulsive and infinitely strong measurement.
Let $a_n$ and $|a_n \rangle$ be the eigenvalues and eigenvectors of 
$\hat A$, respectively 
(for simplicity, we assume a discrete index $n$).
According to Eqs. (\ref{STOCEQ}), 
after a  very short time $t$ and for $\kappa$ so large that $\kappa t \gg 1$
the initial wave function $|\psi(0) \rangle$ transforms into a state
$| \psi_{\xi^{(0)}}(t) \rangle$ which depends on $\xi^{(0)}$, the value of 
the stochastic process at time $t=0$, and has $\hat A$-representation 
\begin{equation}
\langle a_n | \psi_{\xi^{(0)}}(t) \rangle \simeq
\exp \left[ - \kappa t \left( a_n - a(0) - {\xi^{(0)} \over 2 \sqrt{\kappa}}
\right)^2 \right] 
\exp \left( { {\xi^{(0)}}^2 t  \over 4 } \right)
\langle a_n | \psi(0) \rangle, 
\label{PSIT}
\end{equation}
where $a(0)= \langle \psi(0)| \hat{A} |\psi(0) \rangle$.
Provided that $a(0) \neq a_n$ for all $n$, in the limit $t \to 0$ and 
$\kappa t \to \infty$ Eq. (\ref{PSIT}) implies an instantaneous collapse
into the eigenstate of $\hat A$ with eigenvalue closest to
$a(0) + \xi^{(0)}/\sqrt{\kappa}$.
By averaging over all possible values of $\xi^{(0)}$, we get  
\begin{eqnarray} 
\overline{ 
\langle \psi_{\xi^{(0)}}(t)| \hat{A} |\psi_{\xi^{(0)}}(t) \rangle}
&=& \int d\xi^{(0)} \sqrt{t \over 2\pi} 
\exp\left( -{t {\xi^{(0)}}^2  \over 2}\right)
~ \sum_n ~a_n~ 
\left| \langle a_n | \psi_{\xi^{(0)}}(t) \rangle \right|^2
\nonumber \\ &=& 
\sum_n ~a_n~ 
\left| \langle a_n | \psi(0) \rangle \right|^2.
\end{eqnarray}
This is the result expected on the basis of the von Neumann postulate:
in a measurement of $\hat A$ at time $t=0$,
the probability that the state $|\psi(0)\rangle$ collapses into 
the eigenstate $|a_n \rangle$ is  
$\left| \langle a_n | \psi(0) \rangle \right|^2$.
Analogously, Eq. (\ref{LINDBEQ}) provides an 
instantaneous diagonalization of the density matrix in the 
$\hat A$-representation. 

The basic Eqs. (\ref{STOCEQ}) and (\ref{LINDBEQ}) 
can be obtained by considering an explicit model of measurement apparatus. 
We let the measured system, described by the classical Hamiltonian $H(p,q)$, 
to interact with an infinite set of degrees of freedom $(P_n,Q_n)$ via a 
harmonic potential. 
The total system is closed and the Hamiltonian is given by $H_{tot} = H + H_{env}$, where
\begin{equation} 
H_{env} = 
\sum_{n} \left( 
\frac{ P_{n}^2 }{ 2 M} +
\frac{M \omega_{n}^2}{2} \left[ Q_{n} - 
\lambda A(p,q) \right]^2 \right).
\label{HENV}
\end{equation}
The constant $\lambda$ is a transduction factor.
By choosing initial conditions corresponding to the environment variables
at thermal equilibrium with temperature $T$ around the instantaneous state 
of the measured system, the reduced density matrix of the measured system
$\rho(q_1,q_2,t)$ can be exactly evaluated \cite{MQM}.
Moreover, for a proper continuous distribution of the frequencies $\omega_n$,
Eq. (\ref{LINDBEQ}) is recovered in the high temperature limit.
Equation (\ref{STOCEQ}) is then obtained via a decomposition of the 
two-particle Green function associated to $\rho(q_1,q_2,t)$ into a couple
of single-particle Green functions \cite{MQM}. 
The phenomenological constant $\kappa$ is expressed in terms of the 
parameters of the model. 

The high temperature limit necessary for obtaining Eqs. (\ref{STOCEQ}) and 
(\ref{LINDBEQ}) from the above mentioned model has a definite physical meaning.
In order to avoid paradoxical features, the measurement apparatus has to
be classical with respect to an external observer \cite{CINI} and 
in our model this is obtained for $k_B T \gg \hbar \omega_n$. 

In the case of measurements of position, in Ref. \cite{POP} it has been shown 
that Eqs. (\ref{STOCEQ}) and (\ref{LINDBEQ}) provide a correct behavior
also in the controversial case of measurements on macroscopic systems.
Indeed, classical behavior is always established in the macroscopic limit
due to an instantaneous convergence into properly defined coherent states.

\acknowledgements
It is a pleasure to thank Roberto Onofrio for the long standing and fruitful
collaboration on this topic.

\end{document}